\documentclass[
reprint,preprintnumbers,
nofootinbib,
amsmath,amssymb,
aps,
prd
]{revtex4-1}
\pdfoutput=1
\usepackage{graphicx}
\usepackage[utf8]{inputenc}
\usepackage{flushend}
\usepackage{dcolumn}
\usepackage{bm}
\usepackage{balance}
\usepackage[normalem]{ulem}
\usepackage[colorlinks = true,
            linkcolor = blue,
            urlcolor  = blue,
            citecolor = blue,
            anchorcolor = blue]{hyperref}
\usepackage{verbatim,subfigure}
\usepackage{color,ulem}
\usepackage[english]{babel}
\usepackage{MnSymbol,wasysym}
\input Starburst.fd
\newcommand*\initfamily{\usefont{U}{Starburst}{xl}{n}}\initfamily 

\newcommand{\beq}{\begin{eqnarray}}
\newcommand{\eeq}{\end{eqnarray}}
\usepackage{amsmath}
\usepackage{tikz}
\usepackage{hyperref}
\usetikzlibrary{decorations.pathmorphing}
\usetikzlibrary{shapes.misc}
\tikzset{cross/.style={cross out, draw=black, minimum size=8*(#1-\pgflinewidth), inner sep=0pt, outer sep=0pt},
cross/.default={1pt}}
\usetikzlibrary{patterns,math}
\definecolor{applegreen}{rgb}{0.55, 0.71, 0.0}
\newcommand{\dd}{\mathrm{d}}
\usepackage{contour,xparse,xcolor}
\ExplSyntaxOn
\NewDocumentCommand{\HS}{m}
 {
  \seq_set_split:Nnn \l_tmpa_seq { ~ } { #1 }
  \seq_map_inline:Nn \l_tmpa_seq { \contour{green}{##1} ~ } \unskip
 }
\ExplSyntaxOff

\definecolor{OrangeYellowBlend}{rgb}{1, 0.65, 0}

\begin{document}

\preprint{\texttt{LCTP-24-21, IFT-UAM/CSIC-24-170}}

\title{Krylov Complexity in Mixed Phase Space}

\author{Kyoung-Bum Huh,$^{1}$ Hyun-Sik Jeong,$^{2}$ Leopoldo A. Pando Zayas$^{3,4}$ and Juan F. Pedraza$^{2}$ \vspace{1mm}}

\affiliation{$^1$Wilczek Quantum Center, School of Physics and Astronomy, Shanghai Jiao Tong University, Shanghai 200240, China}
\affiliation{$^2$Instituto de F\'isica Te\'orica UAM/CSIC, Calle Nicol\'as Cabrera 13-15, 28049 Madrid, Spain}
\affiliation{$^3$Leinweber Center for Theoretical Physics, University of Michigan, Ann Arbor, MI 48109, USA}
\affiliation{$^4$The Abdus Salam International Centre for Theoretical Physics, 34014 Trieste, Italy}

%
\begin{abstract}
We investigate the Krylov complexity of thermofield double states in systems with mixed phase space, uncovering a direct correlation with the Brody distribution, which interpolates between Poisson and Wigner statistics. Our analysis spans two-dimensional random matrix models featuring (I) GOE-Poisson and (II) GUE-Poisson transitions and extends to higher-dimensional cases, including a stringy matrix model (GOE-Poisson) and the mass-deformed SYK model (GUE-Poisson). Krylov complexity consistently emerges as a reliable marker of quantum chaos, displaying a characteristic peak in the chaotic regime that gradually diminishes as the Brody parameter approaches zero, signaling a shift toward integrability. These results establish Krylov complexity as a powerful diagnostic of quantum chaos and highlight its interplay with eigenvalue statistics in mixed phase systems.
\end{abstract}

\maketitle
%
\noindent\textbf{Introduction.}
Quantum chaos is a ubiquitous phenomenon with far-reaching implications across physics. It governs the process of thermalization, dictates the complex dynamics of many-body systems, and is central to advancements in information theory, particularly in understanding black holes. While the principles of classical chaos are well-established and extensively studied \cite{hilborn2000chaos}, the characterization of quantum chaos ---especially in the many-body regime--- remains a challenging and less developed field, demanding deeper exploration.

A foundational principle in quantum chaos is the Bohigas-Giannoni-Schmit (BGS) conjecture~\cite{Bohigas:1983er,Bohigas:1984aa,Guhr:1997ve,Bohigas}, which posits that quantum systems with classically chaotic dynamics exhibit energy spectra consistent with random matrix theory (RMT). Hallmarks of RMT, such as level repulsion and spectral rigidity, serve as definitive signatures of late-time quantum chaos~\cite{Bohigas:1983er,Berry1985-mx,Muller:2004nb}. In contrast, early-time quantum chaos manifests as the exponential growth of specific observables, notably out-of-time-order correlators (OTOCs)~\cite{larkin1969quasiclassical,berman1978condition}, characterized by a quantum Lyapunov exponent bounded by $\lambda_L \leq 2\pi k_B T/\hbar$~\cite{Maldacena_2016}. While progress has been made in exploring these regimes, the connection between early- and late-time quantum chaos remains an open question, promising deeper insights into the interplay of chaos, thermalization, and the fundamental limits of quantum dynamics.

Recently, intriguing connections between maximal chaos and black hole physics have emerged~\cite{Shenker:2013pqa,Shenker:2014cwa,PhysRevD.94.126010,Cotler:2016fpe,deBoer:2017xdk,Stanford:2019vob}. At the heart of these insights is the early-time saturation of the `chaos bound'~\cite{Maldacena_2016}, which provides a powerful framework for comparing black holes and quantum many-body systems as some of the most efficient information scramblers in nature. This bound has become crucial for identifying quantum systems with potential holographic duals~\cite{Perlmutter:2016pkf}, highlighting the profound connection between quantum dynamics and gravitational phenomena.

Most physical systems are, however, neither maximally chaotic nor fully integrable; instead, they exhibit a mixed phase behavior where different sectors may exhibit level repulsion and level clustering simultaneously. This highlights the urgent need to refine our understanding of quantum chaos indicators and explore their connections with other diagnostics of mixed systems. Krylov complexity~\cite{Parker:2018yvk,Balasubramanian:2022tpr}, a key tool for assessing quantum chaos, has already shown potential for linking early- and late-time signatures. Although its primary focus has been on characterizing chaotic systems, extending its application to mixed phase systems is a natural next step, which we will investigate in this Letter.

Krylov complexity has been studied across a wide range of quantum chaotic systems, including RMT~\cite{Balasubramanian:2022tpr,Balasubramanian:2023kwd,Tang:2023ocr,Caputa:2024vrn,Bhattacharjee:2024yxj,Jeong:2024oao}, quantum billiards~\cite{Hashimoto:2023swv,Camargo:2023eev,Balasubramanian:2024ghv}, quantum spin chains~\cite{Rabinovici:2021qqt,Scialchi:2023bmw,Gill:2023umm,Bhattacharya:2023xjx,Camargo:2024deu,Scialchi:2024zvq}, and various versions of the Sachdev-Ye-Kitaev (SYK) model~\cite{Rabinovici:2020ryf,Bhattacharjee:2022ave,Hornedal:2022pkc,Erdmenger:2023wjg,Chapman:2024pdw,Baggioli:2024wbz}. It has also been explored in diverse contexts such as topological and quantum phase transitions~\cite{Caputa:2022eye,Afrasiar:2022efk,Caputa:2022yju,Pal:2023yik}, quantum batteries~\cite{Kim:2021okd}, high-energy quantum chromodynamics~\cite{Caputa:2024xkp}, bosonic systems modeling ultra-cold atoms~\cite{Bhattacharyya:2023dhp}, saddle-dominated scrambling~\cite{Bhattacharjee:2022vlt,Huh:2023jxt}, and open quantum systems~\cite{Bhattacharya:2022gbz,Bhattacharjee:2022lzy,Mohan:2023btr,Bhattacharya:2023zqt,Bhattacharjee:2023uwx,Carolan:2024wov,Bhattacharyya:2023grv}. A comprehensive review can be found in~\cite{Nandy:2024htc}.

Although Krylov complexity was originally developed to track the growth of operators~\cite{Parker:2018yvk}, our focus here is on the version that monitors the spread of a time-evolving quantum state~\cite{Balasubramanian:2022tpr}. For time-evolved thermofield double (TFD) states in quantum chaotic systems, Krylov complexity exhibits a distinct four-phase pattern: an initial growth, a peak, a decline, and a plateau. As shown in~\cite{Baggioli:2024wbz}, the peak is a universal feature of quantum chaotic many-body systems and is absent in integrable systems. In this Letter, we extend the analysis of Krylov complexity to TFD states in systems with mixed phase space, where both integrable and chaotic regions coexist. We find a direct correlation between the peak and the Brody distribution of eigenvalues~\cite{Brody1973,Brody:1981aa}, which smoothly interpolates between Poisson and Wigner-Dyson spectra.

%
\vspace{0.1cm}
\noindent\textbf{Preliminaries.}
To evaluate Krylov complexity, the Krylov basis $\{|K_n \rangle\}$ is constructed using the Lanczos algorithm~\cite{Lanczos:1950zz,RecursionBook}. This procedure generates Lanczos coefficients $\{a_n,\,b_n\}$, which capture the system's dynamical properties. These coefficients correspond to the entries in the tridiagonal matrix representation of the Hamiltonian within the Krylov basis, given by:
\begin{align}\label{}
    H|K_n \rangle = a_n | K_n \rangle + b_{n+1} | K_{n+1} \rangle + b_n | K_{n-1} \rangle \,.
\end{align}
The Krylov wave functions $\psi_n(t)$ then evolve according to the recursive differential equation:
\begin{align}\label{DES}
    i \, \partial_t \psi_n(t) = a_n \psi_n(t) + b_{n+1} \psi_{n+1}(t) + b_n \psi_{n-1}(t) \,.
\end{align}
This equation represents the Schrödinger equation in Krylov space for the Hamiltonian $H$, with the time-evolved state given by $|\psi(t) \rangle = \sum_n \psi_n(t) | K_n \rangle$.

Krylov complexity is defined as
\begin{align}\label{eq:Krylov complexity}
    C(t) = \sum_{n} n \, |\psi_n(t)|^2 \,.
\end{align}
It quantifies the average spread of the time-evolving state within the Krylov basis, effectively measuring the dispersion of the wave function over time. In quantum chaotic systems, Krylov complexity evolves through a distinct pattern: a ramp, a peak, a decline, and a plateau~\cite{Balasubramanian:2022tpr}. This structure plays a crucial role in distinguishing chaotic dynamics from integrable ones; in particular, the early-time peak serves as a hallmark of chaos~\cite{Baggioli:2024wbz}, and is absent in integrable systems.

To investigate the transition between chaotic and integrable phases, we introduce the order parameter
\begin{align}\label{eq:diff_peak}
    \Delta C = C(t=t_{\text{peak}}) - C(t\rightarrow\infty) \,,
\end{align}
referred to as the Krylov Complexity Peak (KCP)~\cite{Baggioli:2024wbz}. Here, $C(t=t_\text{peak})$ is the peak value of Krylov complexity, and $C(t\rightarrow\infty)$ represents its late-time average or plateau. A nonzero KCP ($\Delta C \neq 0$) indicates a chaotic system, while $\Delta C = 0$ corresponds to an integrable system.

Following~\cite{Balasubramanian:2022tpr}, we adopt the TFD state as the initial state for the dynamics described in \eqref{DES}, expressed as
\begin{align}\label{eq:TFD state}
    |\psi(0) \rangle = \frac{1}{\sqrt{Z(\beta)}}\sum_n e^{-\frac{\beta E_n}{2}} | n\rangle \otimes | n \rangle \,.
\end{align}
We focus on the maximally entangled case ($\beta=0$), where chaotic features are most pronounced~\cite{Balasubramanian:2022tpr,Huh:2023jxt,Baggioli:2024wbz}. In contrast, increasing $\beta$ tends to weaken these signatures~\cite{Balasubramanian:2022tpr,Huh:2023jxt,Baggioli:2024wbz}. Further, for the maximally entangled state, the late-time saturation of Krylov complexity can be analytically computed as $C(t=\infty) = (L-1)/2$, where $L$ is the size of Hamiltonian~\cite{Erdmenger:2023wjg}. 

\vspace{0.1cm}
\noindent\textbf{Eigenvalue statistics of mixed systems.} Our focus now shifts to considering key examples of systems with mixed dynamics, where the spectrum follows a variant of the Brody distribution~\cite{Brody1973,Brody:1981aa}. Originally introduced to model complex atomic nuclei, this distribution interpolates between the Poisson distribution and the GOE distribution in RMT. A slight generalization is provided by the following probability density function~\cite{Jafarizadeh:2012aa,Sabri:2012opl}:
\begin{equation}\label{GBD}
P(s) =  (b+1) \, c_b \, (\alpha \, s^b + \bar{\alpha}\, s^{b+1}) \, \text{exp}\left(-c_b \, s^{b+1} \right)  \,,
\end{equation}
known as the generalized Brody distribution. Here,
\begin{align}\label{}
\begin{split}
\!\!\!\alpha = 1-\frac{
\left[\frac{\Gamma\left(\frac{b+2}{b+1}\right)}{c_b^\frac{1}{b+1}}\right]^2
-\frac{\Gamma\left(\frac{b+2}{b+1}\right)}{c_b^\frac{1}{b+1}}}{\left[\frac{\Gamma\left(\frac{b+2}{b+1}\right)}{c_b^\frac{1}{b+1}}\right]^2
-\frac{\Gamma\left(\frac{b+3}{b+1}\right)}{c_b^\frac{2}{b+1}}} \,, \,\,\,\,\,
\bar{\alpha} = \frac{
\frac{\Gamma\left(\frac{b+2}{b+1}\right)}{c_b^\frac{1}{b+1}}
-1}{\left[\frac{\Gamma\left(\frac{b+2}{b+1}\right)}{c_b^\frac{1}{b+1}}\right]^2
-\frac{\Gamma\left(\frac{b+3}{b+1}\right)}{c_b^\frac{2}{b+1}}} \,,
\end{split}
\end{align}
and 
\begin{equation}\label{GBDsub}
\!\!\!\!\!c_b = \left[\Gamma\left(\frac{b+2}{b+1}\right)\right]^{\,\epsilon (b+1)}\!\!\!\!\!\!\!\!\!, \quad 
\epsilon = 
\begin{cases}
 +1 \,\, \text{for Poisson $\leftrightarrow$ GOE,} \\
 -1 \,\, \text{for Poisson $\leftrightarrow$ GUE,}
\end{cases}
\end{equation}
where $\epsilon$ takes values based on the symmetry class. When $\epsilon = +1$, the distribution in \eqref{GBD} reduces to the original Brody distribution~\cite{Brody1973,Brody:1981aa}, interpolating between the Poisson ($b=0$) and GOE ($b=1$) distributions. Conversely, for $\epsilon=-1$, \eqref{GBD} interpolates between the Poisson ($b=0$) and GUE ($b=1$) distributions.
\begin{figure*}[]
\centering
\begin{minipage}{0.45\textwidth}
        \centering
        \includegraphics[width=\textwidth]{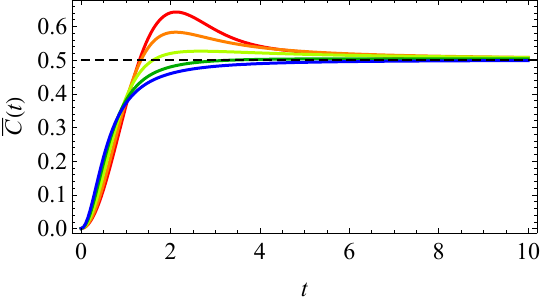}
\end{minipage}
\quad
\begin{minipage}{0.45\textwidth}
        \centering
        \includegraphics[width=\textwidth]{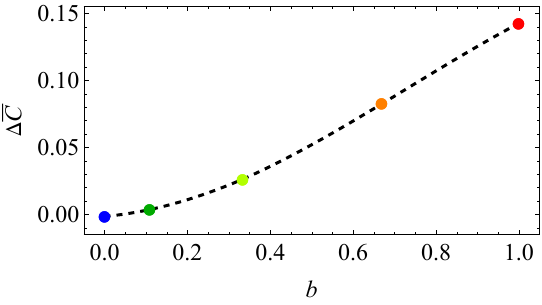}
\end{minipage}
\vspace{-0.3cm}
\caption{\textbf{Left panel:} Averaged Krylov complexity for a matrix model interpolating between GOE and Poisson statistics \eqref{ACT1}, with $\sigma=1$ and $d=1,\, 1.2,\, 1.5,\, 1.8,\, 2$ (red,\, orange,\, yellow,\, green,\, blue). The dashed line represents the late-time plateau. \textbf{Right panel:} KCP $\Delta \bar{C}$ vs. Brody parameter $b$. The colored dots correspond to the same data shown in the left panel.}\label{TWODIMFIG}
\end{figure*}
\begin{figure*}[]
\centering
\begin{minipage}{0.45\textwidth}
        \centering
        \includegraphics[width=\textwidth]{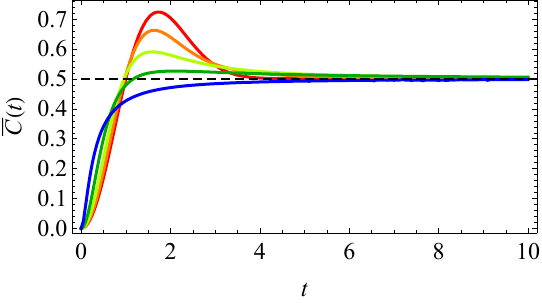}
\end{minipage}
\quad
\begin{minipage}{0.45\textwidth}
        \centering
        \includegraphics[width=\textwidth]{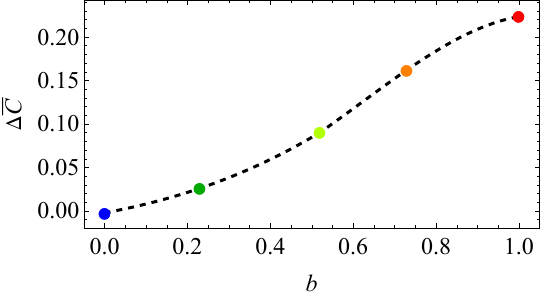}
\end{minipage}
\vspace{-0.3cm}
\caption{\textbf{Left panel:} Averaged Krylov complexity for a matrix model interpolating between GUE and Poisson statistics \eqref{ACTGUE}, with $\sigma=1$ and $d=1,\, 1.2,\, 1.5,\, 2,\, 3$ (red,\, orange,\, yellow,\, green,\, blue). \textbf{Right panel:} KCP $\Delta \bar{C}$ defined in \eqref{eq:diff_peak} vs. Brody parameter $b$. The colored dots correspond to the same data shown in the left panel.}\label{TWODIMFIG2}
\end{figure*}

\vspace{0.1cm}
\noindent\textbf{Primitive $2 \times 2$ random matrix models.}
We first examine the $2 \times 2$ matrix model introduced in~\cite{Nieminen_2017} to build intuition about the role of Krylov complexity in characterizing mixed phases. Its Hamiltonian is given by
\begin{align}\label{ACT1}
    H = 
\begin{pmatrix}
|\mathcal{A}|^d & \frac{1}{2}|\mathcal{B}|^d \\
\frac{1}{2}|\mathcal{B}|^d & 0 
\end{pmatrix} \,,
\end{align}
where $\mathcal{A}$ and $\mathcal{B}$ are real numbers drawn from a Gaussian distribution with zero mean and variance $\sigma$, and $d$ satisfies $1 \leq d \leq 2$. The eigenvalues of $H$ are given by
\begin{align}\label{}
E_{\pm} = \frac{1}{2} \left[ |\mathcal{A}|^d \pm \sqrt{|\mathcal{A}|^{2d} + |\mathcal{B}|^{2d}} \right] \,,
\end{align}
which leads to the level spacing
$s = E_{+} - E_{-}$. The probability density for the spacing, $P(s)$, is derived as follows
\begin{align}\label{ee1}
\begin{split}
P &= \frac{2}{\pi \sigma^2} \int_{0}^{\infty} \dd \mathcal{A}\,\dd \mathcal{B}\,\, \text{exp}\left(-\frac{\mathcal{A}^2+\mathcal{B}^2}{2\sigma^2}\right)    \,\\
&= \frac{2}{\pi d^2 \sigma^2} \int_{0}^{\infty} \dd s \int_{0}^{\frac{\pi}{2}} \dd \phi \,\, s^{\frac{2}{d}-1} \cos^{\frac{1}{d}-1}\phi \, \sin^{\frac{1}{d}-1} \phi  \\ 
&\quad \times \text{exp}\left( - \frac{s^{\frac{2}{d}} \left(\cos^{\frac{2}{d}} \phi + \sin^{\frac{2}{d}} \phi \right)}{2\sigma^2} \right) \\
& \equiv \int_{0}^{\infty} \dd s \, P(s)\,,
\end{split}
\end{align}
where the Jacobian determinant was used in the second equality with $
\mathcal{A}^d = s \cos(\phi)$ and $\mathcal{B}^d = s \sin(\phi)$. As noted in \cite{Nieminen_2017}, the distribution $P(s)$ in \eqref{ee1} closely follows a Brody distribution \eqref{GBD}, with the Brody parameter $b$ related to $d$ in \eqref{ACT1} by $b = (2-d)/d$. Thus, $d=1$ ($b=1$) corresponds to the GOE distribution, while $d=2$ ($b=0$) represents the Poisson distribution. 

Notably, by slightly modifying the above matrix model, we can change the symmetry class and obtain a distribution that transitions between GUE and Poisson. To this end, we consider the following Hamiltonian
\begin{align}\label{ACTGUE}
    H = 
\begin{pmatrix}
|\mathcal{A}|^d & \frac{1}{2} |\mathcal{B}|^d + \frac{i}{2}|\mathcal{C}|^d  \\
\frac{1}{2} |\mathcal{B}|^d - \frac{i}{2}|\mathcal{C}|^d & 0 
\end{pmatrix} \,,
\end{align}
where $\mathcal{A}$, $\mathcal{B}$, and $\mathcal{C}$ are real numbers drawn from a Gaussian distribution with zero mean and variance $\sigma$. In this case, the eigenvalues are given by
\begin{align}\label{}
E_{\pm} = \frac{1}{2} \left[ |\mathcal{A}|^d \pm \sqrt{|\mathcal{A}|^{2d} + |\mathcal{B}|^{2d}+ |\mathcal{C}|^{2d}} \right] \,,
\end{align}
and probability density $P(s)$ follows from
\begin{align}\label{ee2}
\begin{split}
P &= \frac{2\sqrt{2}}{\pi^{\frac{3}{2}} \sigma^{3}} \int_{0}^{\infty} \dd \mathcal{A}\,\dd \mathcal{B}\, \dd \mathcal{C}\,\, \text{exp}\left(-\frac{\mathcal{A}^2+\mathcal{B}^2+\mathcal{C}^2}{2\sigma^2}\right)    \,\\
& \equiv \int_{0}^{\infty} \dd s \, P(s)\,,
\end{split}
\end{align}
where $\mathcal{A}^d = s \sin(\theta) \sin(\phi)$, $\mathcal{B}^d = s \sin(\theta) \cos(\phi)$ and $\mathcal{C}^d = s \cos(\theta)$. Comparing with the generalized Brody distribution \eqref{GBD} for the GUE scenario (where $\epsilon=-1$), we can find a phenomenological relation between the Brody parameter $b$ and $d$, which can be expressed as $b = (3-d)/2d$. Thus, $d=1$ ($b=1$) corresponds to the GUE distribution, while $d=3$ ($b=0$) represents the Poisson distribution. 

With these models established, we now turn to calculating Krylov complexity. For a general system with a two-dimensional Hilbert space in the TFD state, the analytical expression for Krylov complexity is \cite{Caputa:2024vrn}:
\begin{align}\label{}
C_{2\times2} =  \sin^2\left(\frac{s}{2} t \right) \,.
\end{align}
For a random ensemble characterized by the probability density $P(s)$, the remaining task is to compute the averaged Krylov complexity, $\bar{C}$, expressed as
\begin{align}\label{ee3}
\bar{C} \equiv  \int_{0}^{\infty} C_{2\times2} \, P(s) \dd s \,.
\end{align}
In the left panel of Fig.~\ref{TWODIMFIG}, we show numerical results for $\bar{C}$ using $P(s)$ from \eqref{ee1} for $d\in[1,2]$. The peak is maximal at $d=1$ (GOE limit) and diminishes as $d\to2$ (Poisson limit). The right panel recasts this behavior in terms of the Brody parameter $b$, where the KCP defined in \eqref{eq:diff_peak} decreases as $b\to0$. Qualitative similar behaviors are observed for random matrix models interpolating between GUE and Poisson ensembles, as detailed in Fig.~\ref{TWODIMFIG2}. In this case we use $P(s)$ from \eqref{ee2} for $d\in[1,3]$. The peak is maximal at $d=1$ (GUE limit) and diminishes as $d\to3$ (Poisson limit). These results highlight Krylov complexity as a probe of quantum chaos in the mixed phase, consistent with the Brody distribution analysis. 
\begin{figure*}[ht]
\centering
\begin{minipage}{0.45\textwidth}
        \includegraphics[width=\textwidth]{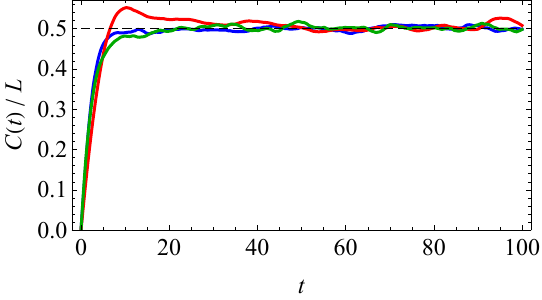}
\end{minipage}
\quad
\begin{minipage}{0.45\textwidth}
        \includegraphics[width=\textwidth]{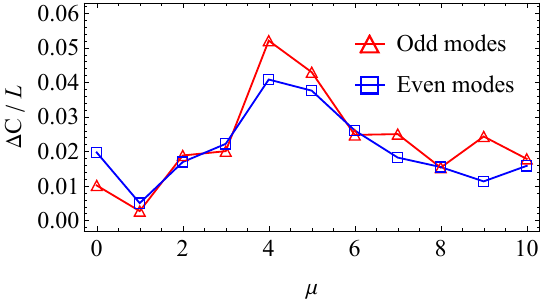}
\end{minipage}
\vspace{-0.3cm}
\caption{\textbf{Left panel:} The Krylov complexity of the odd sector of the stringy matrix model \eqref{BMN} for $\mu=0,\, 4,\, 10$ (blue,\, red,\, green). Similar results are observed in the even sector, but are omitted for brevity. \textbf{Right panel:} KCP $\Delta C$ vs. mass parameter $\mu$. 
The characteristic peak is most pronounced around $\mu=4$, aligning with the expectations from the level statistics. 
}\label{KCPFIGspread}
\end{figure*}
\begin{figure*}[ht]
\centering
\begin{minipage}{0.45\textwidth}
        \centering
        \includegraphics[width=\textwidth]{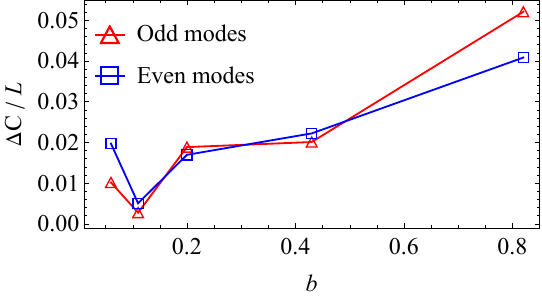}
\end{minipage}
\quad
\begin{minipage}{0.45\textwidth}
        \centering
        \includegraphics[width=\textwidth]{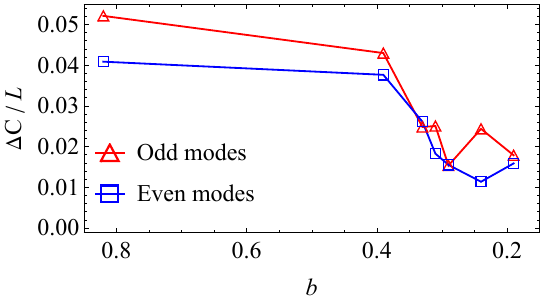}
\end{minipage}
\vspace{-0.3cm}
\caption{\textbf{Left panel}: KCP $\Delta C$ vs. Brody parameter $b$ from $\mu=0$ ($b\approx0$) to $\mu=4$ ($b\approx0.8$). \textbf{Right panel}: KCP $\Delta C$ vs. Brody parameter $b$ from $\mu=4$ ($b\approx0.8$) to $\mu=10$ ($b\approx0$).}\label{DCvsb}
\end{figure*}

\vspace{0.1cm}
\noindent\textbf{Stringy matrix model.}
We now shift our attention to representative examples with higher-dimensional Hilbert spaces, beginning with a truncated version of the stringy BMN matrix model~\cite{Berenstein:2002jq,Banks:1996vh,Amore:2024ihm}. Its Hamiltonian is given by
\begin{equation}\label{BMN}
    H = \frac{p_x^2}{2}+\frac{p_y^2}{2} + \frac{\mu^2 x^2}{8} +\frac{x^4}{2}+x^2y^2 +\frac{\mu^2 y^2}{2} -\mu y^3+\frac{y^4}{2}.
\end{equation}
It describes two coupled non-linear oscillators that model the configuration of two fuzzy spheres, governed by the mass parameter $\mu$. Recent studies of this system~\cite{Amore:2024ihm} have examined its level statistics, identifying a mixed phase space characterized by a Brody distribution.

The Hamiltonian \eqref{BMN} exhibits a $x \rightarrow -x$ symmetry, allowing eigenstates to be classified by even or odd parity with respect to the $x$-coordinate. In our calculations, we used 2,000 eigenvalues for each symmetry sector. As noted in~\cite{Amore:2024ihm}, we find that for both very small and large $\mu$, the Brody parameter $b$ approaches zero. In contrast, for moderate $\mu$, $b$ attains its maximum value of approximately $b\approx0.8$ around $\mu\approx4$.  This indicates a transition in the stringy matrix model \eqref{BMN} from Poisson-like behavior (small $\mu$, $b\approx0$) to chaotic behavior ($\mu\approx4$, $b\approx0.8$), and back to Poisson-like behavior at larger $\mu$ ($b\approx0$). Detailed analysis of this transition in the level statistics is provided in the Supplemental Material~\cite{L121902sup}. 

%
In Fig.~\ref{KCPFIGspread}, we show the normalized Krylov complexity $C(t)/L$ for $\mu=0, 4, 10$. Our analysis reveals two key observations:
(I) A prominent peak appears around $\mu\approx4$, vanishing for sufficiently small or large $\mu$, in agreement with level statistics.
(II) Unlike other chaos diagnostics, such as OTOCs and the spectral form factor (SFF) explored in \cite{Amore:2024ihm}, Krylov complexity correctly identifies integrability at $\mu=0$ by the absence of a peak. This shows that KCP is insensitive to instabilities from flat directions, making it a more reliable probe of quantum chaos. Moreover, the agreement between level spacing and Krylov complexity suggests a direct correlation between the KCP and the Brody parameter $b$, as demonstrated in our earlier two-dimensional toy models. Indeed, we find that this relationship holds for the stringy matrix model, as illustrated in Fig.~\ref{DCvsb}. These findings reinforce Krylov complexity as a valuable indicator of quantum chaos, consistent with the level spacing analysis in mixed systems.
\begin{figure*}[]
\centering
\begin{minipage}{0.45\textwidth}
        \centering
        \includegraphics[width=\textwidth]{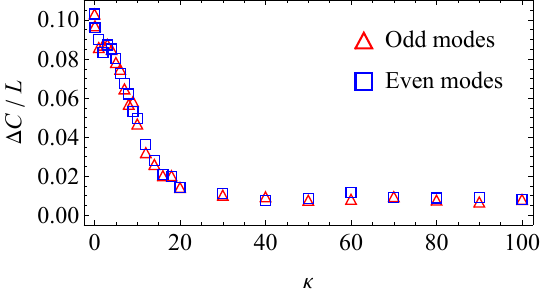}
\end{minipage}
\quad
\begin{minipage}{0.45\textwidth}
        \centering
       \includegraphics[width=\textwidth]{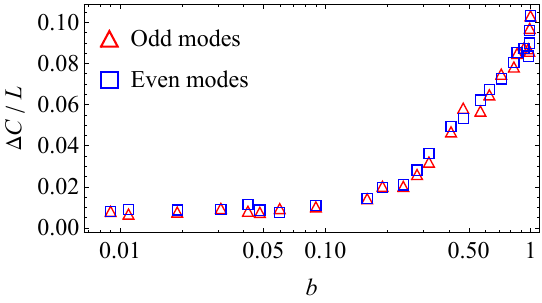}
\end{minipage}
\vspace{-0.3cm}
\caption{\textbf{Left panel}: KCP $\Delta C$ vs. mass parameter $\kappa$ in the deformed SYK model. \textbf{Right panel}: KCP $\Delta C$ vs. Brody parameter $b$. The smooth monotonic dependence between the two quantities demonstrate the reliability of the KCP to characterize mixed phase systems. }\label{MDSYKPHASEFIG}
\end{figure*}

\vspace{0.1cm}
\noindent\textbf{Mass-deformed SYK model.}
A second example of a system with a higher-dimensional Hilbert space and a mixed phase space ---this time interpolating between the Poisson and GUE distributions--- is the mass-deformed SYK model~\cite{Song_2017,Eberlein_2017,Garcia-Garcia:2017bkg}. Its Hamiltonian is given by
\begin{align}\label{GSYK}
    H = \frac{1}{4!}\sum^N_{i,j,k,l=1}\, J_{ijkl}\, \chi_i\, \chi_j\, \chi_k\, \chi_l\, + \frac{i}{2!}\, \sum^N_{i,j=1}\,\kappa_{ij}\, \chi_i\,\chi_j \,,
\end{align}
where $\chi_i$ are Majorana fermions obeying the anticommutation relations $\{\chi_i, \chi_j\} = \delta_{ij}$, while the coupling constants $J_{ijkl}$ and $\kappa_{ij}$ are Gaussian random variables with zero mean and standard deviations $\sqrt{6}J/N^{3/2}$ and $\kappa/\sqrt{N}$, respectively. The Hamiltonian \eqref{GSYK} features a conserved charge parity operator, splitting the system into parity-even and parity-odd sectors~\cite{You:2017aa,Cotler:2016fpe,Krishnan:2017lra}.

The undeformed SYK model (without the mass deformation, i.e., the $\kappa_{ij}$ term) is maximally chaotic, saturating the chaos bound~\cite{Maldacena_2016}, while the purely quadratic Hamiltonian corresponds to an integrable system. Depending on the number of Majorana fermions, $N$, the SYK model's spectrum aligns with different random matrix ensembles ---GUE, GOE, or GSE~\cite{You:2016ldz}. Here, we set $N = 26$, a standard choice in SYK studies, placing the system in the GUE ensemble in its chaotic phase. Increasing the mass deformation parameter $\kappa$ drives a transition from chaos to integrability, a phenomenon captured by level statistics and OTOCs~\cite{Garcia-Garcia:2017bkg,Nosaka2018,Kim:2020mho,Garcia-Garcia:2020dzm,Lunkin:2020tbq,Nandy:2022hcm,Menzler:2024atb,C_ceres_2022,Orman:2024mpw,Garc_a_Garc_a_2021,Caceres:2023yoj,Garcia-Garcia:2023jlu} as well as by Krylov complexity~\cite{Baggioli:2024wbz}. Notably, \cite{Baggioli:2024wbz} established the KCP as a robust diagnostic of this phase structure, complementing traditional chaos indicators.

The system's level statistics is, in fact, well described by a generalized Brody distribution. Our analysis shows that as $\kappa$ increases, the Brody parameter $b$ decreases monotonically, marking a smooth transition from a GUE distribution at $\kappa=0$ (where $b=1$) to a Poisson distribution at $\kappa=70$ (where $b \approx 0.02$). This transition is illustrated in detail in the Supplemental Material~\cite{L121902sup}.

A detailed study of Krylov complexity in the mass-deformed SYK model was conducted in~\cite{Baggioli:2024wbz}; here, we focus on its correlation with the Brody distribution. The left panel of Fig.~\ref{MDSYKPHASEFIG} shows the KCP as a function of the mass deformation parameter $\kappa$ for both parity-even and parity-odd sectors, revealing a decrease in KCP at larger $\kappa$, consistent with the findings of~\cite{Baggioli:2024wbz}. The right panel of Fig.~\ref{MDSYKPHASEFIG} illustrates the relationship between the KCP and the Brody parameter $b$, showing a clear monotonic behavior. These results strengthen the case for the KCP as a reliable indicator of the chaos-to-integrability transition, aligning with level statistics and demonstrating a direct correlation with the Brody distribution.

%
\vspace{0.1cm}
\noindent\textbf{Discussion.}
In this work, we investigated Krylov complexity in systems with mixed phase spaces, encompassing both chaotic and integrable regimes. Focusing on TFD states, we analyzed systems with energy level spacing distributions described by a generalized Brody distribution, where the Brody parameter $b$ interpolates between Poisson statistics ($b=0$) and random matrix ensembles ($b=1$). We demonstrated that Krylov complexity effectively captured the transition from quantum chaos to integrability, with a distinct peak in the chaotic regime that diminished as integrability increased, in precise correlation with the Brody parameter. Our examples included both few-body and many-body systems, with deterministic and probabilistic spectra, further reinforcing the robustness of our conclusions.

This correlation is significant given the discrepancies observed among various quantum chaos indicators. For instance, in~\cite{Hashimoto:2020xfr}, the authors examined the inverted harmonic oscillator (an integrable system that is classically unstable) and found exponential growth of the OTOC, a behavior typically associated with chaotic systems. More recently, studies on the stringy matrix model~\cite{Amore:2024ihm} reported similar discrepancies for the SFF and OTOC correlators vis-\`a-vis the Brody distribution, which were similarly attributed to classical instabilities of the model. Our investigation is thus particularly valuable, as it provides complementary insights into time regimes where OTOCs alone may offer an incomplete or misleading characterization of chaos and integrability.

The correlation between Krylov complexity and the Brody parameter provides crucial quantitative insights into the dynamics of mixed phase systems, offering a deeper understanding of chaotic-integrable transitions and shedding light on potential connections between chaos, quantum information, and black hole physics.

%
\vspace{0.1cm}
\noindent \emph{Acknowledgements.}s
HSJ, and JFP are supported by the Spanish MINECO ‘Centro de Excelencia Severo Ochoa' program under grant SEV-2012-0249, the ‘Atracci\'on de Talento’ program (Comunidad de Madrid) grant 2020-T1/TIC-20495, the Spanish Research Agency via grants CEX2020-001007-S and PID2021-123017NB-I00, funded by MCIN/AEI/10.13039/501100011033, and ERDF `A way of making Europe.' LPZ is partially supported by the U.S. Department of Energy under grant DE-SC0007859. KBH acknowledges the support of the Foreign Young Scholars Research Fund Project (Grant No.22Z033100604). 

\vspace{0.1cm}
\noindent All authors contributed equally to this paper and should be considered
as co-first authors.
\newpage
%
\bibliographystyle{apsrev4-1}
\bibliography{Refs}

%
\onecolumngrid 
\section*{Supplemental material}
In this Supplemental Material, we provide a more detailed characterization of the level spacing statistics for our two examples with higher-dimensional Hilbert spaces. 
\begin{figure}[]
 \centering
      \subfigure[\,\,Even modes, $b=0.06$]
     {\includegraphics[width=4.3cm]{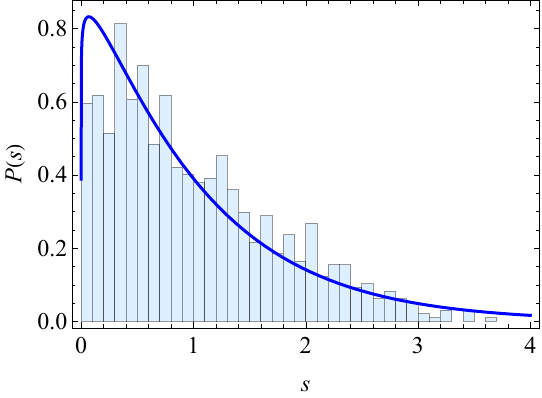}\label{}}
    \qquad
      \subfigure[\,\,Odd modes, $b=0.08$]
     {\includegraphics[width=4.3cm]{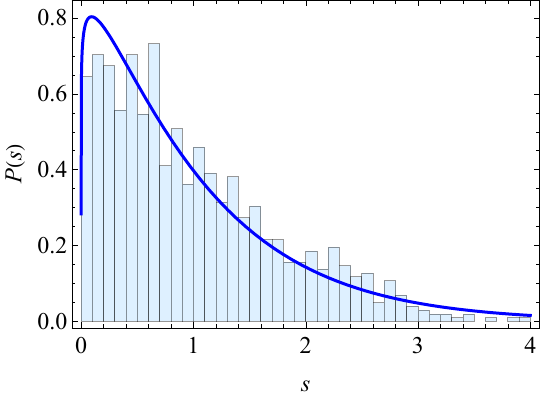}\label{}} 
     \vspace{-0.3cm}
     \caption{Level spacing of stringy matrix model \eqref{BMN} for $\mu=0$.}\label{LSFIG1} 
 
      \subfigure[\,\,Even modes, $b=0.20$]
     {\includegraphics[width=4.3cm]{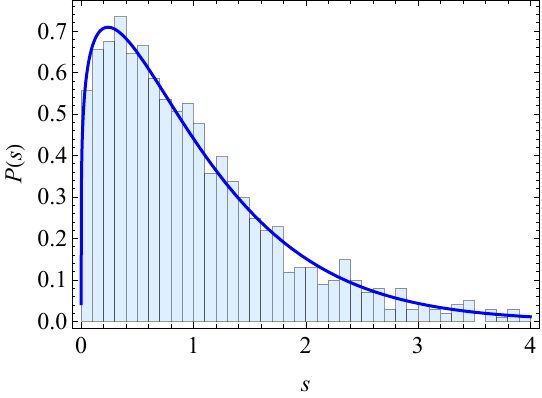}\label{}}
     \qquad
      \subfigure[\,\,Odd modes, $b=0.18$]
     {\includegraphics[width=4.3cm]{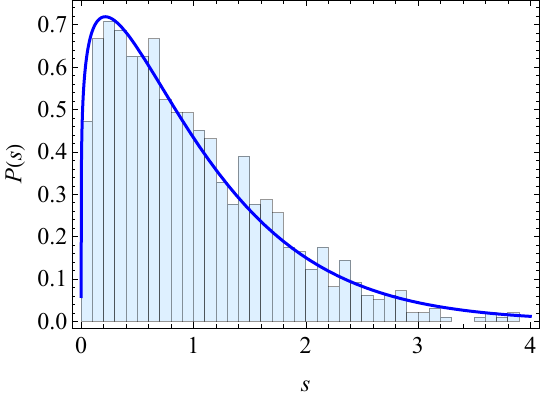}\label{}} 
     \vspace{-0.3cm}
     \caption{Level spacing of stringy matrix model \eqref{BMN} for $\mu=2$}\label{LSFIG2} 

      \subfigure[\,\,Even modes, $b=0.82$]
     {\includegraphics[width=4.3cm]{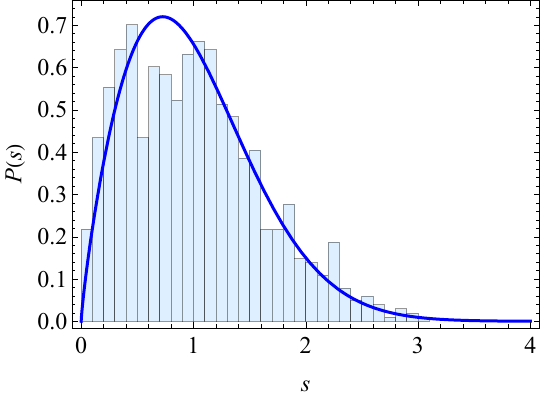}\label{}}
     \qquad
      \subfigure[\,\,Odd modes, $b=0.8$]
     {\includegraphics[width=4.3cm]{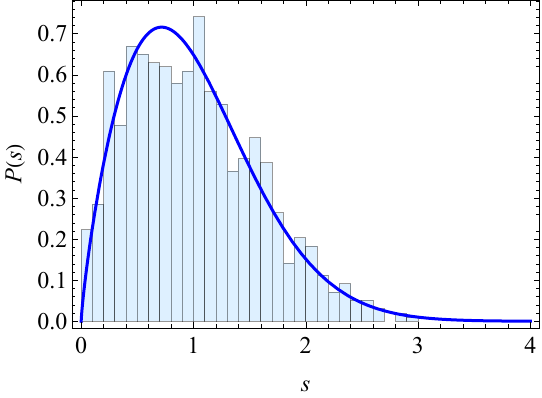}\label{}} 
     \vspace{-0.3cm}
     \caption{Level spacing of stringy matrix model \eqref{BMN} for $\mu=4$.}\label{LSFIG3} 
     
      \subfigure[\,\,Even modes, $b=0.39$]
     {\includegraphics[width=4.3cm]{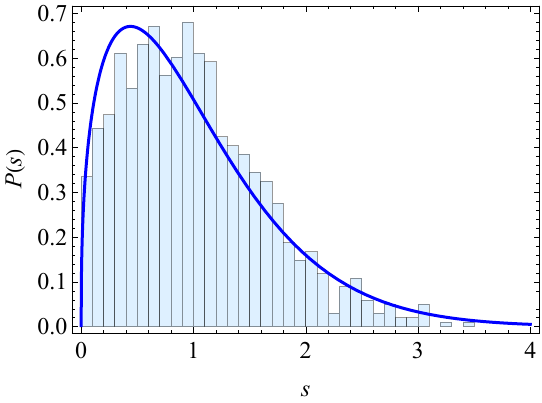}\label{}}
     \qquad
      \subfigure[\,\,Odd modes, $b=0.42$]
     {\includegraphics[width=4.3cm]{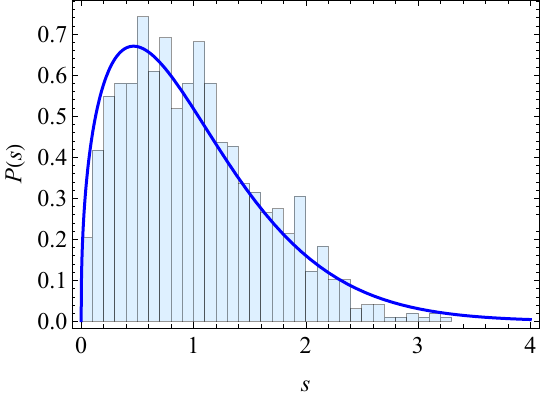}\label{}} 
     \vspace{-0.3cm}
     \caption{Level spacing of stringy matrix model \eqref{BMN} for $\mu=5$.}\label{LSFIG4} 

      \subfigure[\,\,Even modes, $b=0.19$]
     {\includegraphics[width=4.3cm]{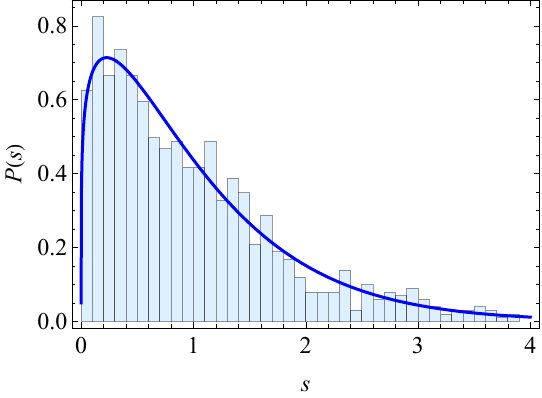}\label{}}
     \qquad
      \subfigure[\,\,Odd modes, $b=0.21$]
     {\includegraphics[width=4.3cm]{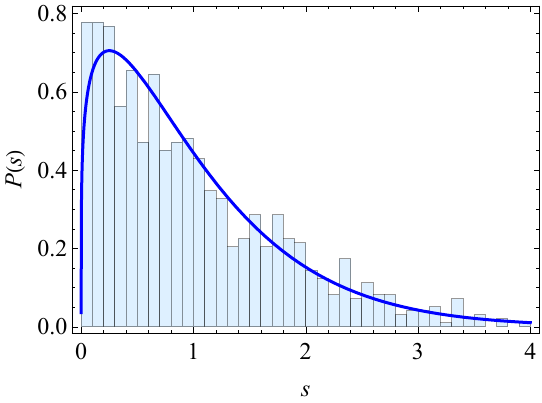}\label{}} 
     \vspace{-0.3cm}
     \caption{Level spacing of stringy matrix model \eqref{BMN} for $\mu=10$.}\label{LSFIG5} 
\end{figure}

\section{Stringy matrix model}
\begin{figure*}[ht]
\centering
        \centering
        \includegraphics[width=0.45\textwidth]{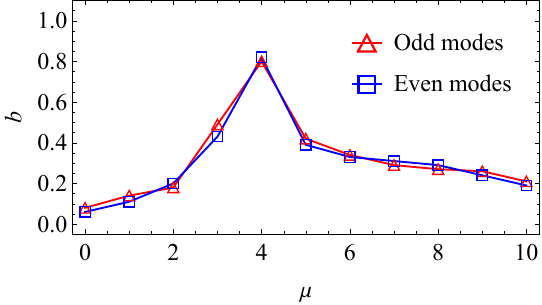}
\vspace{-0.3cm}
\caption{Brody parameter $b$ vs. mass parameter $\mu$ in the stringy matrix model.}\label{BrodySMM}
\end{figure*}

Figs.~\ref{LSFIG1}-\ref{LSFIG5} present the level spacing distributions (histograms) of the stringy matrix model, alongside a comparison with the Brody distribution \eqref{GBD} for the GOE-Poisson case \eqref{GBDsub} (blue line) across various values of $\mu$. Consistent with \cite{Amore:2024ihm}, our results reveal a transition from Poisson-like behavior at small $\mu$ to chaotic behavior at moderate $\mu$, and back to Poisson-like at larger $\mu$. Notably, our finer sampling of $\mu$ identifies the Brody parameter reaching its maximum value, $b\approx0.8$ (indicating closest proximity to the GOE distribution), at $\mu\approx4$, compared to $\mu\approx5$ as reported in \cite{Amore:2024ihm}. The detailed relationship between the mass parameter $\mu$ and the Brody parameter $b$ is shown in Fig.~\ref{BrodySMM}.

\section{Mass-deformed SYK model}
\begin{figure}[]
 \centering
      \subfigure[\,\,Even modes, $b=1$]
     {\includegraphics[width=4.3cm]{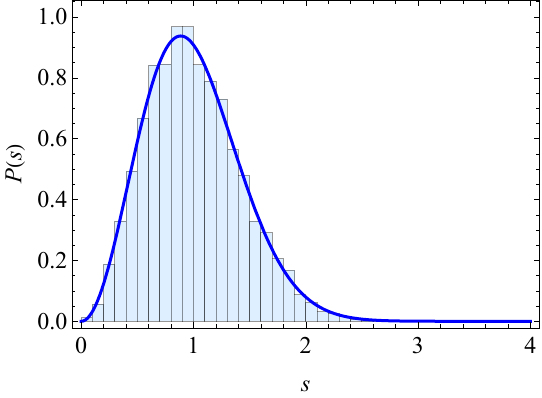}\label{}}
     \qquad
      \subfigure[\,\,Odd modes, $b=1$]
     {\includegraphics[width=4.3cm]{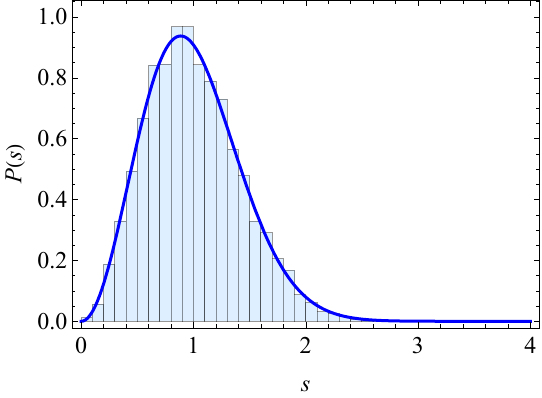}\label{}} 
          \vspace{-0.3cm}
     \caption{Level spacing of mass-deformed SYK model \eqref{GSYK} for $\kappa=0$.}\label{SYKLSFIG1} 
 
      \subfigure[\,\,Even modes, $b=0.83$]
     {\includegraphics[width=4.3cm]{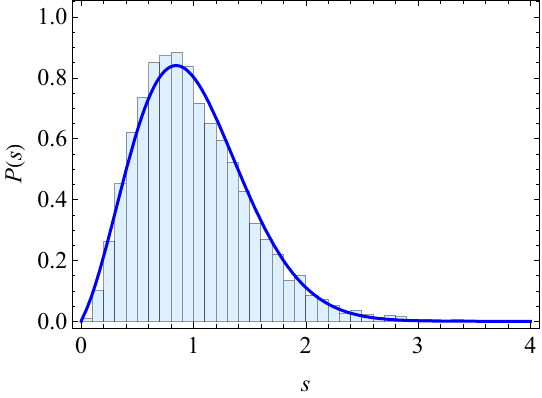}\label{}}
     \qquad
      \subfigure[\,\,Odd modes, $b=0.84$]
     {\includegraphics[width=4.3cm]{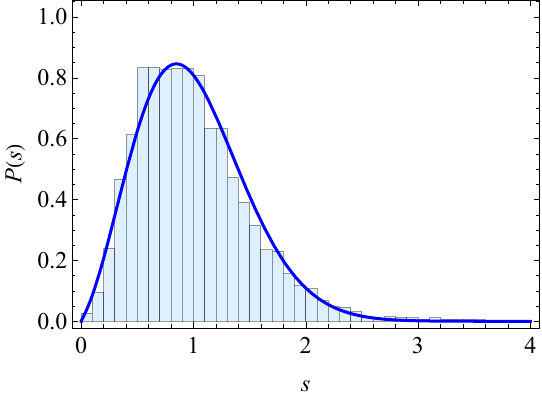}\label{}} 
          \vspace{-0.3cm}
     \caption{Level spacing of mass-deformed SYK model \eqref{GSYK} for $\kappa=5$}\label{SYKLSFIG2} 

      \subfigure[\,\,Even modes, $b=0.06$]
     {\includegraphics[width=4.3cm]{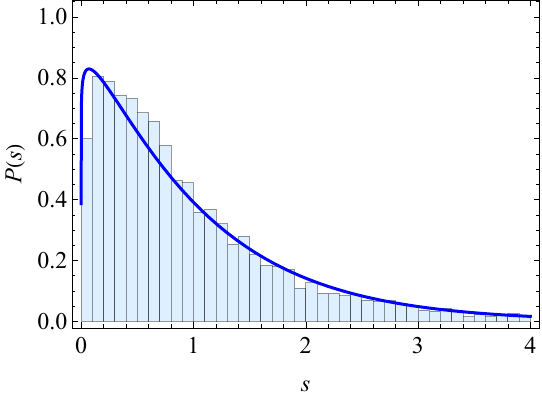}\label{}}
     \qquad
      \subfigure[\,\,Odd modes, $b=0.06$]
     {\includegraphics[width=4.3cm]{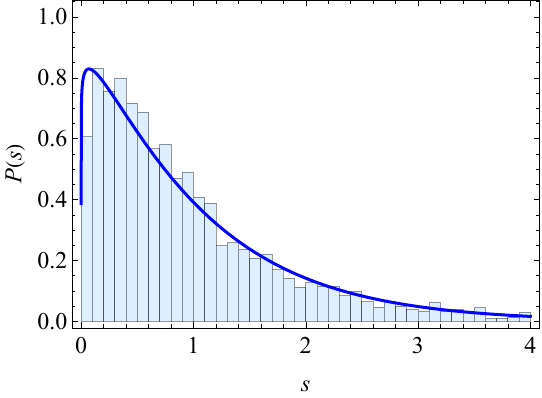}\label{}} 
          \vspace{-0.3cm}
     \caption{Level spacing of mass-deformed SYK model \eqref{GSYK} for $\kappa=40$.}\label{SYKLSFIG3} 
     
      \subfigure[\,\,Even modes, $b=0.03$]
     {\includegraphics[width=4.3cm]{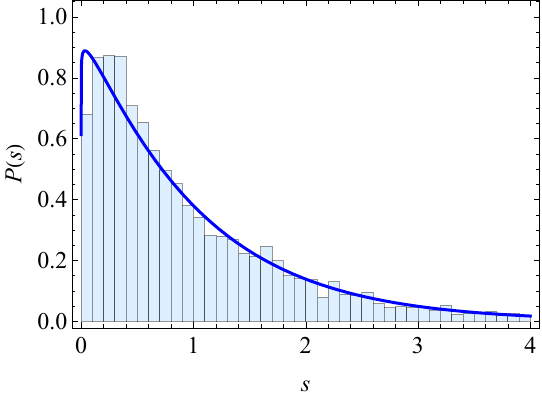}\label{}}
     \qquad
      \subfigure[\,\,Odd modes, $b=0.02$]
     {\includegraphics[width=4.3cm]{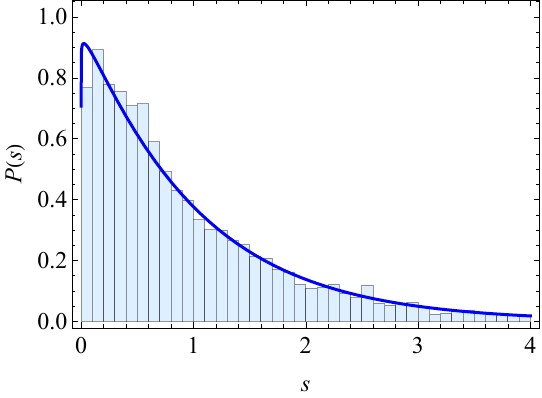}\label{}} 
     \vspace{-0.3cm}
     \caption{Level spacing of mass-deformed SYK model \eqref{GSYK} for $\kappa=70$.}\label{SYKLSFIG4} 
\end{figure}

\begin{figure*}[]
        \centering
        \includegraphics[width=0.45\textwidth]{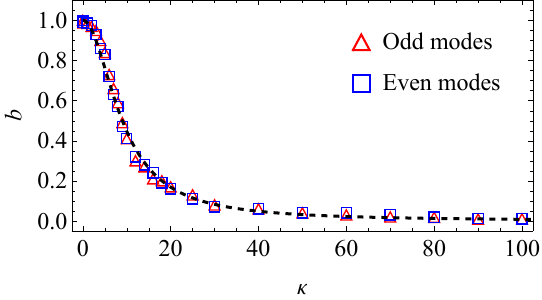}
\textbf{\vspace{-0.3cm}}
\caption{Brody parameter $b$ vs. mass parameter $\kappa$, in the mass deformed SYK model.}\label{BrodySYK}
\end{figure*}

In Figs.~\ref{SYKLSFIG1}-\ref{SYKLSFIG4}, we present the level spacing distributions (histograms) of the mass-deformed SYK model, compared to the Brody distribution \eqref{GBD} for the GUE-Poisson case \eqref{GBDsub} (blue line), across different values of $\kappa$. As $\kappa$ increases, the system transitions from a GUE distribution at $\kappa=0$ (where $b=1$) to a Poisson distribution at $\kappa=70$ (where $b \approx 0.02$). The precise relationship between the mass parameter $\kappa$ and the Brody parameter $b$ is depicted in Fig.~\ref{BrodySYK}.

\end{document}